# Global Sea Level Stabilization-Sand Dune Fixation: A Solar-powered Sahara Seawater Textile Pipeline

Viorel Badescu•
Candida Oancea Institute of Solar Energy
Faculty of Mechanical Engineering
Polytechnic University of Bucharest
Spl. Independentei 313
Bucharest 79590
ROMANIA
e-mail: badescu@theta.termo.pub.ro

Richard B. Cathcart
Geographos
1300 West Olive Avenue, Suite M
Burbank, California 91506
USA

Alexander A. Bolonkin
C & R
1310 Avenue R, 6-F
Brooklyn, New York 11229
USA

**Abstract**

Could anthropogenic saturation with pumped seawater of the porous ground of active sand dune fields in major deserts (e.g., the westernmost Sahara) cause a beneficial reduction of global sea level? Seawater extraction from the ocean, and its deposition on deserted sand dune fields in Mauritania and elsewhere via a Solar-powered Seawater Textile Pipeline (SSTP) can thwart the postulated future global sea level. Thus, Macro-engineering offers an additional "cure" for anticipated coastal change, driven by global sea level rise, that could supplement, or substitute for (1) stabilizing the shoreline with costly defensive public works (armoring macroprojects) and (2) permanent retreat from the existing shoreline (real and capital property abandonment). We propose Macro-engineering use tactical technologies that sculpt and vegetate barren near-coast sand dune fields with seawater, seawater that would otherwise, as commonly postulated, enlarge Earth's seascape area! Our Macro-engineering speculation blends eremology with hydrogeology and some hydromancy. We estimate its cost at $1 billion—about 0.01% of the USA's 2007 Gross Domestic Product.

**I. Introduction**

*Circa* 3900 BC—about 1,100 years after global sea level stabilized following the Last Glacial Maximum—civilization commenced with the human use of newly abundant coastal margin resources; *circa* 2300 BC, urban governments commenced construction of monumental infrastructures.[1] Contemporaneous destabilization of global sea level impacts urban governments and infrastructures on present-day coastal margins.[2] For example, some major cities—places such as Lagos, Karachi, Mumbai, Kolkata, Bangkok, Jakarta, Manila and Shanghai—presently find their fresh groundwater reservoirs are being contaminated by saltwater intrusions that may be further aggravated by a postulated greater, yet unproved, future global sea level rise.[3] "Global Warming" appears to be a real existential risk, buts its impact during the 21st Century and beyond could plausibly range from nil to neglible to severe. Modern-day computer models of future atmospheric realities are still extremely simplistic as compared to the actual phenomenon of Earth's air.

---

• Corresponding author.

Worldwide, since ~10.5% of the world's population reside on land that is <100 km from the shoreline at elevations <10 m above sea level (~2.2% of all land), the international and intranational legal implications of coastal zone adjustments instigated by the alleged impending global sea level rise are profound.[4] Hydrogeology may have a central role to play in a macroproject solving any prospective future global sea level rise because that profession already has a burgeoning role in the underground sequestration (injection and storage) of aerial $CO_2$ gas in deep saline aquifers.[5] For example, during this century, BP PLC operates an isolated natural-gas processing plant in the Sahara near In Salah ($27^0$ 12' North Latitude by $2^0$ 28' East Longitude), Algeria.[6]

The macroproject here proposed involves a massive anthropogenic redistribution of Spaceship Earth's seawater cargo. Whatever the cause of a future global sea level rise, whether it is "global warming" or some other phenomenon or aggregation of phenomena, James E. Hansen defines a substantial global sea level rise to mean "a total sea level rise of at least two meters, because that would be sufficient to flood large portions of Bangladesh, the Nile Delta, Florida, and many island nations, causing forced migration of tens to hundreds of millions of people."[7] More than 20 years ago, Walter Stephenson Newman (1895-1978) and Rhodes Whitmore Fairbridge (1914-2006) speculated that *Homo sapiens* could, using Macro-engineering tactical technologies, manage any future global sea level rise by "…diverting sea water into continental depressions" filled to present-day global sea level.[8] By our estimation, the Caspian Sea region could store ~13,000 $km^3$, the Aral Sea region ~1,000 $km^3$, the Qattara Depression ~3,200 $km^3$, the Dead Sea ~1,260 $km^3$, Lake Eyre region ~200 $km^3$ and the Salton Sea region ~400 $km^3$ of seawater. Total global depression storage capacity: ~18,060 $km^3$ and all of the seawater is stored in the open air, subject to solar evaporation. In other words, these seawater storehouses must be constantly replenished at some undetermined financial and energy cost! The area of the Earth's ocean is ~3.62 x $10^8$ $km^2$. If the ocean rose by James E. Hansen's two meters, the volume of that increase would be ~725,000 $km^3$, indicating that removal of ~18,060 $km^3$ is only ~2.5% of the volume that must be shifted to the Earth's land from the ocean's basin in order just to maintain present-day global sea level. Where can the remaining 97.5% (~705,940 $km^3$) be stored, displaced and withheld from the ocean, for an indeterminate time period at a reasonable financial cost?

First, a maximum of 7,200 $km^3$ of freshwater, about 20% of the Earth's total annual river runoff is retained in artificial reservoirs created by anthropic dams.[9] Second, the total natural unused, unsaturated volume of pore space beneath the world's land is huge—to a depth of ~2,000 m, if only unconsolidated sands, sandstones and carbonates are considered, ~25,000,000 $km^3$ exists which might be filled artificially with seawater! "Sands retain most of their original porosity down to a depth of 1 km. Porosities of approximately 48 percent at the surface show little change for the initial 100 m of burial, and then begin to decrease slightly with depth: to 45 percent at 300 m and 37 percent at 1 km."[10] Water of any purity can be forced to accumulate rapidly in the unsaturated void space of porous materials such as sand and sandstone above the natural water table.[11] (Oilfield repressurization with deliberately injected seawater has been used for many years to halt widespread land subsidence caused by oil and natural gas mining.) Any removal of the ocean's excess seawater need only be pumped away at the same rate it is naturally added by the effects of "Global Warming".

Large active sand dune fields are found, generally, in Earth's desert regions—"Hot Deserts" cover ~14.2% of Earth's land.[12] Some eremologists suspect that "global desertification", a persistent decline of ecosystems' benefits for humans in dry areas, is occurring and will increase as the 21st Century unfolds.[13] "Drylands cover about 41% of Earth's land surface and are home to more than 38% of the total global population of 6.5 billion."[14] Here, however, we focus only on certain active sand dune fields located in the northern Africa nation of Mauritania[15] where few people live and work today.[16]

Mauritania's active and widespread sand dune fields are depositional landforms of increasingly better known accumulation history. They contain many cubic kilometers of aeolian-moved sediment and are situated at three major depositional physiographic sites: (1) Akchar; (2) Aouker and at (3) Majabat al Koubra, the farthest inland.

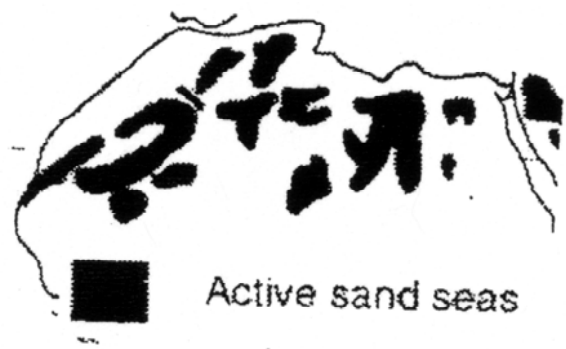

### "Western Africa Desert Dunes"

The active sand dune fields of Mauritania exist in a region that receives <150 mm of precipitation per year. Dune sand is an inert "soil" without any positive characteristics for flora due to its coarse particles and big pore spaces that do not retain water for plant growth, the high permeability (~5000 millidarcy!) and leaching that removes plant nutrient elements, and the wind erodibility of sand.

Aeolian movement of surface sand grains and other dry granular media can be reduced or stopped by application of chemical stabilizers such as "Nano Clay"[17] or "Biochar"[18], installation of fences to trap windblown sand grains, and planted vegetation to prevent sand grain deflation. The present-day and future movement of sand dunes, ground surface erosion of drought-reduced lakes[19], and accumulation of mineral dust clouds, including halite and other salts, challenge Macro-engineering practice as well as seriously affect the lives and life-styles of the people downwind from major active and inactive dune fields.[20] Macro-engineering has, so far, only a few field-tested techniques to geographically fix migrating sand dunes: (1) remove the sand mechanically; (2) disperse the sand by mechanical reshaping and (3) immobilize the sand dunes with planted vegetation[21], fences, trenches, additives and other means. Such macroprojects, of course, involve Nature's further domestication by *Homo sapiens*.[22]

Chad's infamous Bodele Depression[23] is Earth's largest single source of atmospheric mineral dust; Mauritania's eastern drylands and desert is the second largest source of mineral dust in Africa. The mineral dust clouds emanating from northern Africa, triggered when near-surface boundary layer wind speeds are >10 m/sec, quickly change the planet Earth's albedo and can strongly suppress eastern North Atlantic Ocean hurricane formation.[24] In other words, the Sahara's uncontrolled or mitigated mineral dust storms do have major consequences for humans.[25] If mineral dust clouds were eliminated technologically, then northern Africa might become a well-vegetated region of the planet. Summarizing, we propose to foster plant growth—both wild and cultivated—on the sandy dunal surfaces of Mauritania, Chad and Libya by the land's irrigation with pumped seawater and that seawater withdrawal from the world's ocean would also serve to induce a lowering of global sea level to prevent a worldwide "rising sea level crisis" as suggested by James E. Hansen and many other experts.

**II. The Active Sand Dune Field Macro-problem**

James E. Hansen's global climate warming scenario entails a global sea level rise, "…a process that will shift the interface between land and sea, resulting in the inland extension of maritime related flooding and elevated soil salinities".[26] Under natural circumstances, a tenfold reduction in aeolian sand migration can be induced by a mere 3% moisture increase and the deposition of sea-spray halite particles on subaerial seashore dune sand increases the angle of repose of coastal sand dunes.[27] Injection water seepage effects on the lee-side slope of sand dunes generally acts to reduce the angle of repose while water suction acts generally to increase the critical slope (angle of repose).[28] It is nowadays well-known that sand saturated with hypersaline solutions does retain more moisture than sand saturated with moderately and slightly hypersaline solutions.

It is estimated ~10% of Earth's land is already affected by salt deposition. Of ~5,000 food and fiber crops that are cultured by humans, only a few can survive with water that contains >0.5% salt, and most suffer serious yield reductions at ~0.1% salt. Still, the use of saline waters and even seawater for halophytic crop cultivation is an attractive option for farmers in some dryland regions.[29] (The world's first commercial food to be grown entirely on poor soil irrigated by seawater is *Salicornia bigelovii*. The increasing salinization of inland waters[30], and the recent calls for "reversing the flow of water and nutrients from the ocean to the land"[31], combined with the amazing prospects for progressive genomic manipulation of photosynthetic plants, means that saline water will soon have a greater value to humans than it has had in the recent historic past.) Cropping of vast additional segments of the Sahara[32], or the Sahara's near-term future coverage by a Sahara Tent Greenbelt[33], might curtail or even terminate the natural suppression of North Atlantic Ocean tropical cyclones! Sand moves by creep, saltation and aerial suspension. Seawater sprinkled onto the surface of mobile sand dunes would deposit minerals in the space between granular materials, especially after the freshwater is evaporated by natural daytime solar energy—could this surface-deposited material be harvested from the tops of artificially stabilized sand dunes by inexpensive machines? After evaporation, the mineral-rich layer and/or encrustation will contain many useful materials, and its mining, could obviate some current "polluting" mine excavation and processing operations. The irrigation potential for Mauratania using renewable freshwater resources is considered negligible[34]; sand dune seawater sprays, immitating Nature's strand sea spray, are a new form of irrigation but for the single purpose of mining seawater elements![35] In a sense, humans could mine artificial "ores" emplaced during a very short period of time; seawater consists of at least 80 elements. Beach mining is common for obtaining iron, diamonds and materials for concrete's creation.

**III. Mauritania Prototype Macroproject Site**

Mauritania is mostly desert—it is constantly hot, dry and dusty and mostly barren, with flat plains [elevation extremes: lowest place is Sebkhet Te-n-Dghamcha (-5 m), highest place is Kediet Ijill (915 m)] of the Sahara and vast encroaching sand dune field that now threaten to inundate the nation's post-1960 Capital of Nouakchott ($18^0$ 09' North Latitude by $15^0$ 58' West Longitude). The average insolation level for Nouakchott, Mauritania, is approximately 6.55 kWh/m$^2$/day. This Capital could be modernized by the installation of shade-providing photovoltaic "Trees of Paradise", as proposed for the yet-to-be-built facility "Senscity Paradise" Las Vegas, Nevada, in the USA.[36]

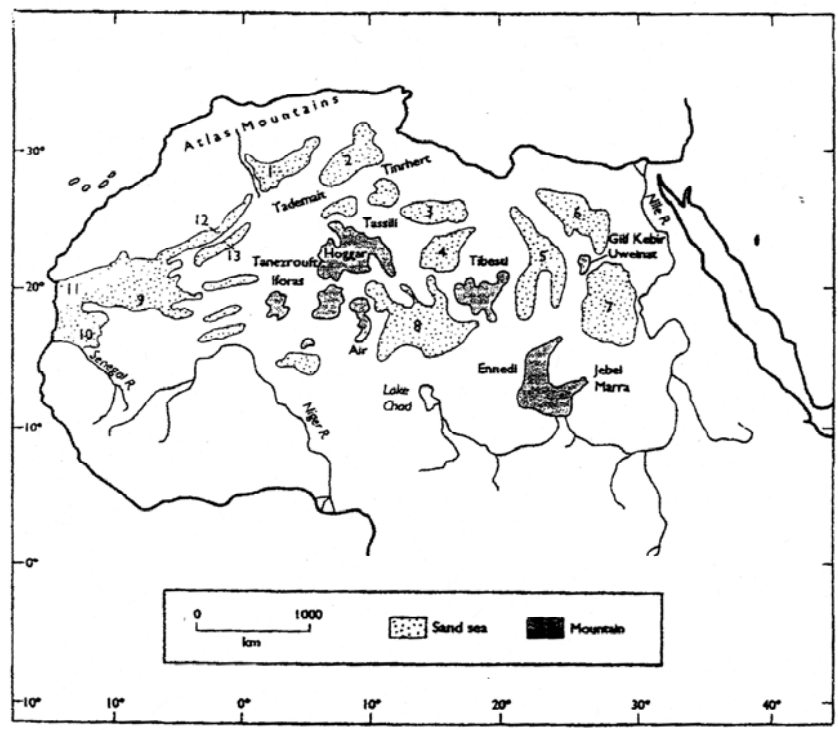

**"Physiographic Features of the Sahara"**

The human population is ~3.1 millions and is distributed very discontiguously over an area of 1.03 million km$^2$; most persons are concentrated in the near-sea level Capital, a seaport since 1987, in the seaport of Nouadhdibou and along the Senegal River in the southern part of Mauritania. Mauritania's territory includes approximately 25% of the Senegal River Basin (~75,500 km$^2$). Mauritania and Senegal are the only two countries in northern Africa with all their agricultural production located within drylands. Freshwater is so valuable a commodity to those nations sharing the Senegal River's surface runoff that only ~3% of the river's outflow reaches the North Atlantic Ocean. Constructed in the Senegal River Delta by 1986, the Diama Dam blocks tidal or storm surge seawater intrusions upriver. Less than 500 km$^2$ of Mauritania's land is irrigated with freshwater! Of a total annual production of 191 million kWh, only 14% of Mauritania's electricity production derives from hydropower while 86% is manufactured from fossil fuel combustion.

**IV. Modern and Improving Solar Energy Technologies**

Flexible solar power assemblies include a flexible photovoltaic device attached to a flexible thermal solar collector.[37] Thin-Film Solar Cells: Next Generation Photovoltaics and Its Applications, edited by Y. Hamakawa and published in 2005 by Springer in New York, is one of the best sources for data.[38] The two-volume book set, Energy from the Desert (Earthscan, 2007), by Kosuke Kurokawa, is a reliable source of direct applicational information. Our macroproject invokes a seawater pipeline-integrated photovoltaic flexible solar power module or membrane that won't ever—or, at least, for a period of 10-15 years—debond under very harsh desert conditions of sunshine and windblown particle abrasion. The pliable photovoltaic coating will need to cover less than half the curved upper-hemisphere of the steel and/or concrete pipe, which act as the support structure for the proposed pliable electricity-generation skin.

Inspired by United States Patent 5,160,214, "Irrigation System and Irrigation Method", issued to Mikio Sakurai and Chikako Sakurai on 3 November 1992, we noticed their map (Figure 1, reproduced as our Figure 3, below) really seemed, fortuitously, to resemble northern Africa and, in particular, Mauritania!

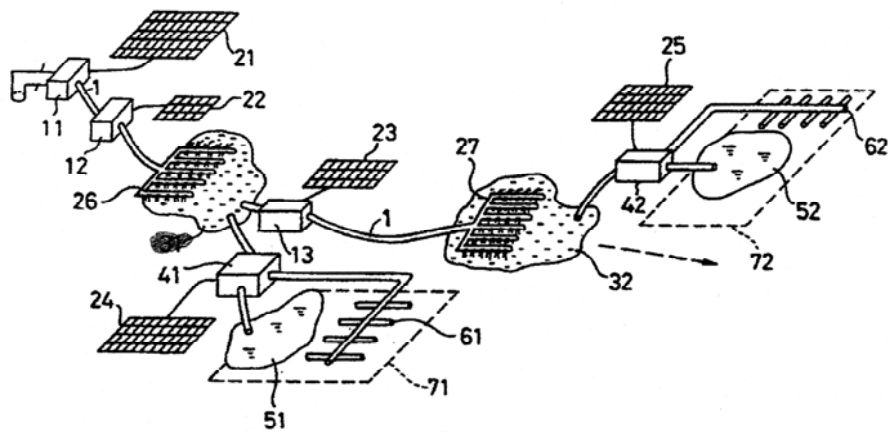

**"Sakurai Patent Figure 1"**

The identity of shape can be compared with Thomas Schluter's Geological Atlas of Africa with Notes on Stratigraphy, Tectonics, Economic Geology, Geohazards and Geosites of Each Country (Springer, 2006). The Sakurai's placement of the inland artificial lake seems even to represent Mauratiania's world-famous 38 kilometer-wide "Richat Structure" ($21^0$ 04' North Latitude by $11^0$ 22' West Longitude), with a central depth elevation of 400 m and surrounding walls nearly 100 m higher. The Richat Structure exposes a flat-lying limestone in the Maur Adrar Desert.[39] The Richat Structure tops out at an elevation of nearly 600 m, yet the center part of the crater-shaped geomorphological feature is ~400 m above present-day sealevel. For maximum use as a seawater pool, it may be necessary to build a small tensioned textile dam on the Richat Structure's southwester edge, where the elevation is only ~414 above sea level.

Roger H. Charlier, during 1991, suggested a route for a freshwater-carrying steel or concrete pipeline starting from the seaport of Nouakchott and heading eastward into the Sahara, finally changing direction to meet Libya.[40] Since Charlier's macroproject proposal entailed the importation of freshwater, and its transportation to the middle of the Sahara, he obviously intended to supplement Libya's Great Man-made River Project (GMRP).[41] The GMRP—it planning started *circa* 1983-84—was built after the 1953 discovery of freshwater aquifers beneath the Sahara. Almost everywhere in Libya, pipes are used to transport groundwater from one region to another without causing ground erosion and reducing the chance of wasteful evaporation. There are already 4,000 km of pipe laid, mainly of 4 m-diameter pre-stressed concrete; ultimate freshwater delivery is expected to reach ~6.0 million $m^3$. (Comparable in length—that is, ~4,000 km-long—is the proposed Trans-Saharian Gas Pipeline will pump natural gas from Nigeria via Niger and Algeria.)

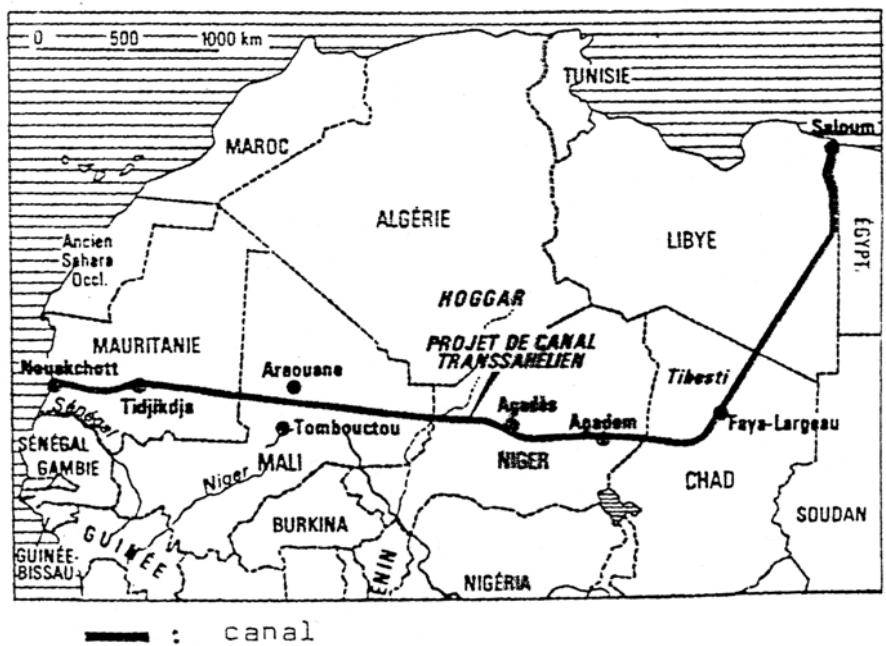

Proposed routing of a trans-Sahara canal.

## "Charlier's proposed routing of trans-Sahara Canal"

We propose to utilize the same route, in part, for the Solar-powered Sahara Seawater Textile Pipeline—that is, from the Capital Nouakchott to Tidjikdya ($18^0$ 27' North Latitude by $11^0$ 27' West Longitude) the distance is ~486 km while we would build a textile pipeline section leaving Tidjikdya and ending near Ouadane ($20^0$ 51' North Latitude by $11^0$ 37' West Longitude), a distance of ~265 km. Such a routing permits use of the Richat Structure as a pooled seawater resource base from which other activities may be carried out per the Sakurai's patented suggestions! Whether the Sakurai's knew it or not, this facility will not contaminate the fresh groundwater held in the "Continental Terminal" formation first discovered in 1931.[42]

Humans will, more and more, harness the Sun's energy. Currently, *Homo sapiens* commands ~13 TW globally. To use 20 TW more than today, all generated by photovoltaic cells that work with only a 10% efficiency, people need to landscape approximately 0.16% of Earth's land. Deserts seem likely to become prominent landscaped features in the near-term future! Of course, if there is a revolutionary technical improvement in flexible solar photovoltaic cell technology, then less land would have to be dedicated to electricity generation.[43]

**IV. The Solar-powered Sahara Seawater Textile Pipeline (SSTP)**

Seawater's speed in the proposed Solar-powered Sahara Seawater Textile Pipeline hermetic tube can be estimated by equation

$$V = \frac{4m}{\pi D^2}, \quad (2)$$

where $V$ is seawater speed, m/s; $m$ is seawater extension, m$^3$/s; $D$ is the tube diameter, m.

Computation is presented in FIGURE 1. The greater tube diameter will promote less seawater flow speed and, thus, will reduce inside seawater losses.

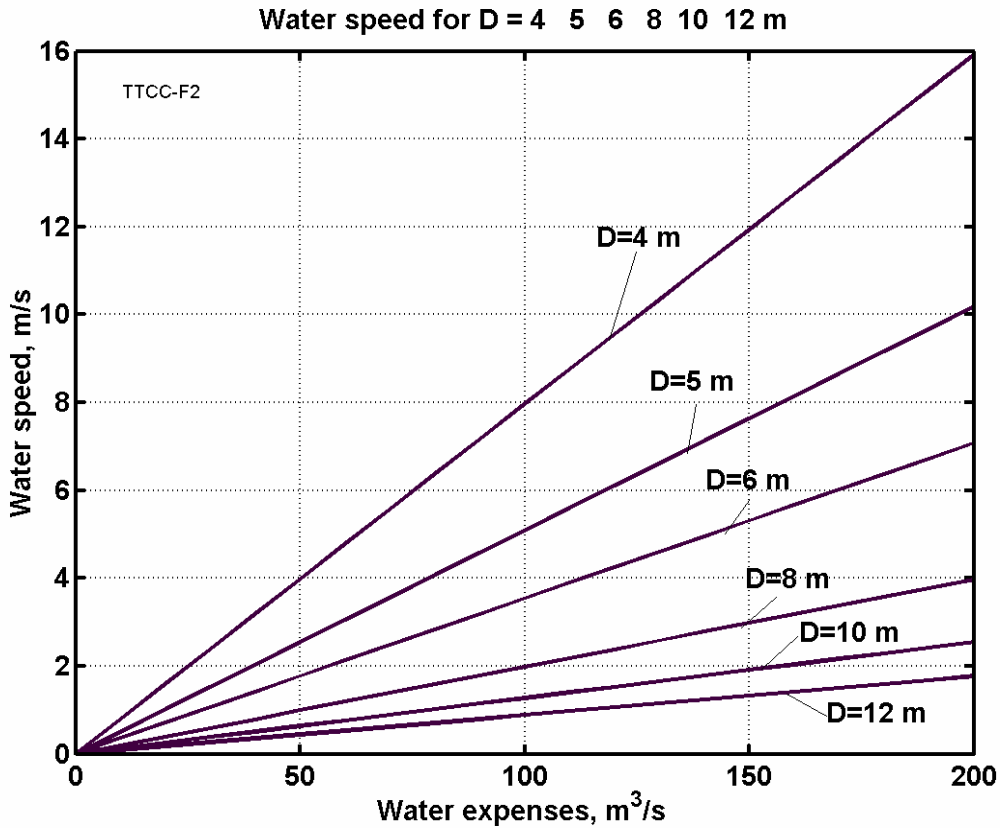

**FIGURE 1.** Seawater speed in tubular SSTP via water expenses for different tube diameters.

The loss of seawater pressure (in m) is computed by equation

$$h = f \frac{L}{2g} \frac{V^2}{D}, \quad (3)$$

where $h$ is loss of seawater pressure, m; $f$ is friction coefficient; $L$ is length of tube, m; $V$ is seawater speed, m/s; $D$ is the tube diameter, m. The seawater friction coefficient is

$$f = \frac{0.25}{\left[\log\left(\dfrac{k}{3.7D} + \dfrac{5.14}{R_e^{0.4}}\right)\right]^2}, \tag{4}$$

where $R_e$ is the applicable Reynolds number, $k$ is the roughness factor. In our singular SSTP case, we assume the use of reinforced concrete or steel pipes/tubes because textile friction coefficient is still iffy. The friction coefficient for both is approximately 0.06. The computation of equation (3) is presented in FIGURE 2. Loss amounts to about 200 m of seawater pressure over a distance of 200 km or 130 m over a distance of 150 km.

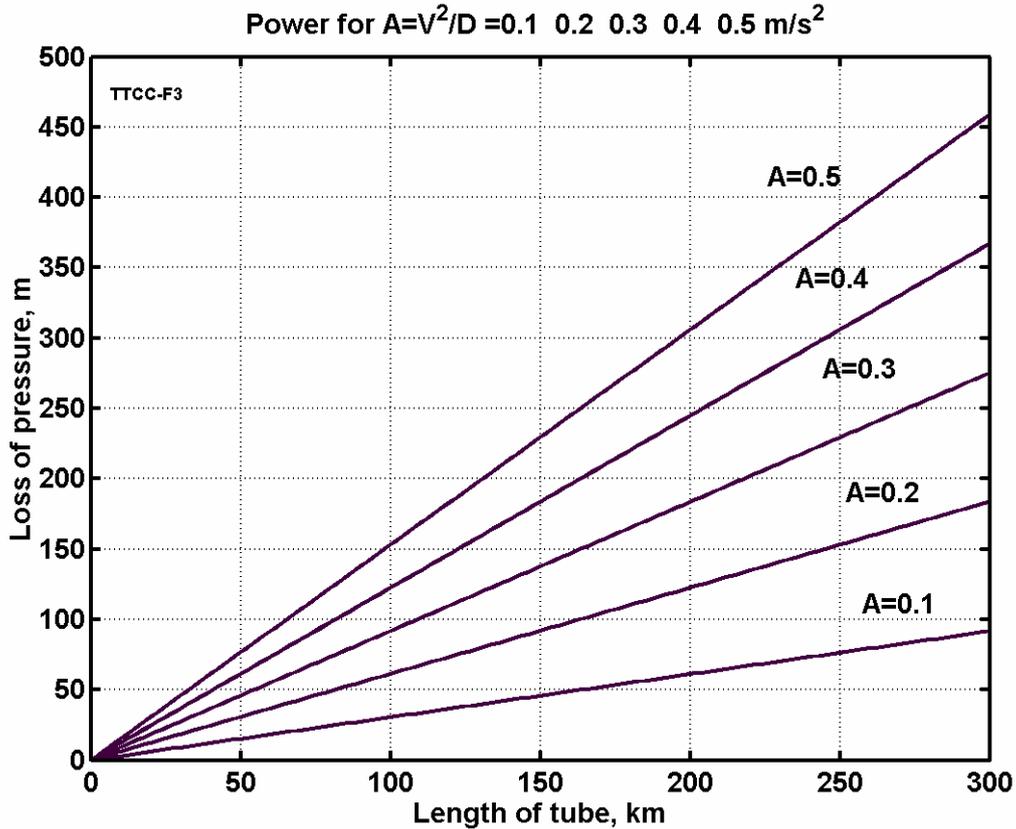

**FIGURE 2.** Loss of the seawater pressure into the closed pipe system via the length of tube for the different ratios $A=V^2/D$, where $V$ is seawater speed, $D$ is hermetic tube diameter. Friction coefficient $f = 0.06$.

The relative loss of the tube's seawater pressure may be estimated by equation

$$\bar{h} = f \frac{V^2}{2gD}, \tag{5}$$

where $\bar{h} = h/L$ is relative loss of seawater pressure, m/km. The computation is presented in FIGURE 3 below.

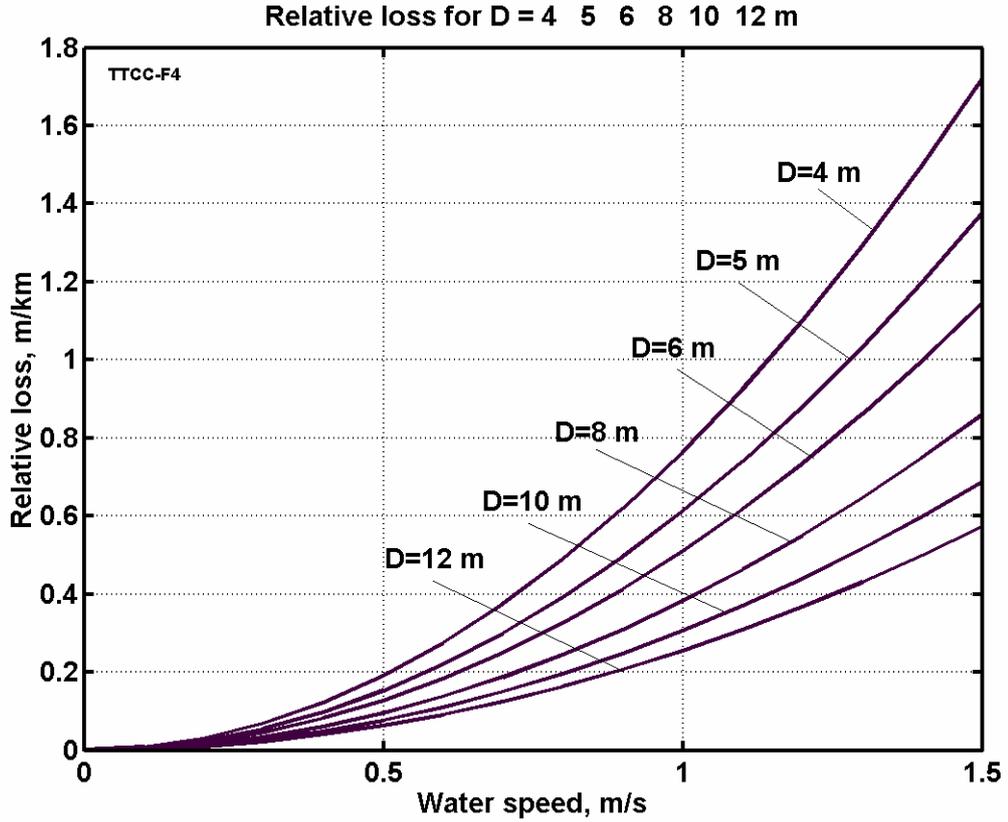

**FIGURE 3.** Relative loss (m/km) of the seawater pressure via the seawater speed for different SSTP diameters. Friction co-efficient $f = 0.06$.

The presumed big hydraulic pipeline (with a large tube diameter) is more expensive but such a tube has the advantage of decreasing greatly the pressure loss and increases significantly the efficiency. The SSTP pipes/tubes may have high interior pressure so they must be composed of steel or something even stronger—perhaps from some 21$^{st}$ Century-invented composite fiber material. Super textiles will become very useful. The absence of metals would, of course, mean the absence of rusting and, perhaps, reduced interior fouling. Such composed material—perhaps stronger than Kevlar—has higher maximum stress (up to 600 kg/mm$^2$, steel has only ~120 kg/mm$^2$) and low specific-density (~1800 kg/m$^3$, steel is ~7900 kg/m$^3$). Due to expected manufacturing efficiencies, it become cheaper quite soon during the 21$^{st}$ Century. Coefficient of safety is 3 to 5. Below, is the equation for computation of the necessary tube wall thickness

$$\delta = \frac{pD}{2\sigma}, \qquad (6)$$

where $\delta$ is tube-wall thickness, m; $p$ is seawater pressure, N/m$^2$; $\sigma$ is safety tensile stress, N/m$^2$. The power needed for pumping seawater may be computed by equation

$$P = \frac{\pi\rho}{4} D^2 V(H + h), \qquad (7)$$

where $P$ is power, W; $\rho$ is seawater density, 1000 kg/m$^3$; D is tube diameter, m; $V$ is water speed into pipe, m/s; $H$ is altitude of final tube end, m; $h$ is loss of seawater friction, m.

*Example*: For the pipe length $L = 100$ km and $D = 4$ m$^2$ and $V = 1.5$ m/s, we have $h = 160$ m, $H = 500$ m; $P = 12.4$ MW.

The requisite area of photovoltaic solar cells may be estimated by equation

$$A = \frac{P}{P_u} = \frac{P}{\eta \eta_s P_s},  \qquad (8)$$

where $A$ is area of photovoltaic cells, m$^2$; $P_s$ is the maximum solar photovoltaic cell power at the Earth's Equator per 1 m$^2$, $P \approx 10^3$ W/m$^2$; $\eta \approx 0.5$ is average efficiency co-efficient of solar radiation in during daylight; $\eta_s \approx$

$0.2 \div 0.45$ is efficiency coefficient of solar cells; $P_u$ is used solar power, W/m$^2$.

*Example*: For $P = 12.4$ MW (see 5, above), $\eta_s \approx 0.3$ the needed solar cell area is 83,000 m$^2$. The solar area of tube $L = 100$ km and $D = 4$m$^2$ is At = 400,000 m$^2$.

***In other words, only ~20% of the SSTP hermetic pipeline's total surface is enough to power electric water pumps to displace seawater inland from the ocean by this tube with a speed V = 1.5 m/s.***

This SSTP installation can pump the amount of seawater:

$$M = \frac{\pi \rho}{4} D^2 V t,  \qquad (9)$$

where $M$ is mass (or volume, m$^3$/time) of seawater, kg/time; $\rho$ is seawater density. kg/m$^3$ (or specific volume m$^3$/time), $t$ is time (seconds, day, year).

*Example*: For tube above ($L = 100$ km, $D = 4$ m) in daylight ($t = 12$ hours $= 43200$ sec) the deliverable seawater productivity will be 814,000 m$^3$/day. Sprinkled onto active sand dunes to make them far less mobile, at some future time, the following chemical substances could be harvested if were desired by Mauritanians (or others):

| ELEMENT | WEIGHT 0/0 | KILOGRAMS |
|---|---|---|
| **Chlorine** | 0.0194 | 16186390 |
| **Sodium** | 0.0108 | 9010980 |
| **Magnesium** | 0.001292 | 1077980.2 |
| **Sulfur** | 0.00091 | 759258.5 |
| **Calcium** | 0.0004 | 333740 |
| **Bromine** | 0.000067 | 55901.45 |
| **Carbon** | 0.000028 | 23361.8 |

**V. Conclusion**

We have shown that an SSTP macroproject sited in Mauritania can be both economic and developmental, with extensive future applicability worldwide. The SSTP's Macro-engineering concept of extraction and long-term storage of excess seawater in threatening active coastal sand dune fields via simple and effective tactical technologies (artificially duplicating, in part, the Earth's natural Hydrologic Cycle) is revolutionary. Monetarily, we suspect SSTP construction would cost 2007 USA $1 billion.

## VI. REFERENCES and ACKNOWLEDGMENTS

**Acknowledgment:** We wish to thank Dr. Stephen Salter in Scotland for his inspirational idea for sand sequestration of excess ocean seawater and Dr. J. Marvin Herndon for his chemistry advice.